\documentclass[]{aa}
\usepackage{graphicx}
\usepackage{txfonts}

\begin{document}

\title{High-precision astrometry on the VLT/FORS1 at time scales of few days
        \thanks{Based on observations made with the European Southern
        Observatory telescopes obtained from the ESO/ST-ECF Science
        Archive Facility}}

\author{P.F.Lazorenko 
        	\inst{1} 
  \and  M.Mayor \inst{2}
  \and  M.Dominik \inst{3} 
     \thanks{Royal Society University Research Fellow}  
  \and  F.Pepe \inst{2}
  \and  D.Segransan \inst{2}     	
    \and  S.Udry \inst{2}
   	}

   \offprints{P.F.Lazorenko}

     \institute{Main Astronomical Observatory,
          National Academy of Sciences of the Ukraine,
             Zabolotnogo 27, 03680 Kyiv-127, Ukraine
          \and  Observatoire de  Geneve, 51 Chemin des Maillettes, 
          1290 Sauverny, Switzerland   
          \and  SUPA, University of St Andrews, School of Physics \&
            Astronomy, North Haugh, St Andrews, KY16 9SS, United Kingdom
              }

   \date{}                                                            

   \abstract{ }
{We investigate the accuracy of astrometric measurements with the VLT/FORS1 camera and
consider potential applications.
}
{The study is 
based on two-epoch (2000 and 2002/2003) frame series of observations
of a selected Galactic Bulge sky region that were
obtained with FORS1 during four consecutive nights each. 
Reductions were carried out with a novel technique that
eliminates atmospheric image motion and does not require a distinction
between targets and reference objects. 
}
{The positional astrometric precision was found to be
limited only by the accuracy of the determination of the star photocentre,
which is typically 200--300~$\mu$as per single measurement for bright unsaturated
stars $B=18-19$.
Several statistical tests have shown that at time-scales of
1--4 nights the residual noise in measured
positions is essentially a white noise  with no systematic
instrumental signature and no significant deviation from a Gaussian
distribution. Some evidence of a good astrometric quality of the VLT
for frames separated by two years has also been found. }
{
Our data show that
the VLT with FORS1/2 cameras can be effectively used for astrometric
observations of planetary microlensing events and other applications
where a high accuracy is required, that is expected to reach
 30--40~$\mu$as
for a series of 50 frames (one hours with $R_{\rm{special}}$ filter). 
}
{}

      \keywords{astrometry   --  atmospheric effects  --
                    instrumentation: high angular resolution --
                    planetary systems}

%  \titlerunning{High-precision astrometry on the VLT/FORS1}
%  \authorrunning{Lazorenko P.F. et al. } 

     \maketitle

\section{Introduction}
The provision of high-precision astrometry (of a few 100~$\mu$as and
below) is much desired, but so far it is not widely available.
Over the recent years, photometric microlensing has proven its feasibility
for not only detecting massive gas giant planets (Bond et al. \cite{Bond};
Udalski
et al. \cite{Udalski}), but even cool rocky/icy planets (Beaulieu et al. 
\cite{Beaulieu}).
A fourth planet, of roughly Neptune-mass, was most recently claimed
by Gould et al. (\cite{Gould}). 
Several further potential detections of planets
by microlensing have however been missed (e.g. Jaroszy\'{n}ski \&
Paczy\'{n}ski \cite{Jaros}) since a proper characterization requires good
data quality and a dense sampling. In fact, it is necessary to overcome
ambiguities and degeneracies of binary lenses themselves (Dominik 
\cite{Dominik99}),
between mass ratio and source size (Gaudi \cite{Gaudi}), 
and between binary-lens
and binary-source systems 
(Gaudi \cite{Gaudi98}), as shown explicitly by Gaudi \& Han
(\cite{Gaudi04}). 
The observation of the shift of the centroid of light composed
of the lens star and the images of the source star with time, so-called
astrometric microlensing (H{\o}g et al. \cite{Hog}; 
Hardy \& Walker \cite{Hardy};
Miyamoto \& Yoshii \cite{Miyamoto}; 
Dominik \& Sahu \cite{DominikS}), provides a two-dimensional
vector, whose absolute value is proportional to the angular Einstein
radius (comprising information about the lens mass and relative
lens-source parallax), in addition to the magnification provided by the
photometric
observations, where the latter only allows to measure the angular Einstein
radius if the finite size of the source star can be assessed in the
light curve. While Safizadeh et al. (\cite{Safizadeh}) 
first discussed the possibility
of using astrometric microlensing for discovering and characterizing
extra-solar planets, Han \& Chunguk (\cite{Han}) found that the planetary
deviation to the astrometric signal expected from a Galactic bulge star
lensed by a foreground star and a surrounding Jupiter-mass planet is
about 10-200~$\mu$as. Rather than requiring a stability of the detector
on time-scales that correspond to the orbital period of planets as for
the detecting of the astrometric shift seen in its host star due to motion
around the common center-of-mass (e.g. 
Quirrenbach et al. \cite{Quirrenbach};
Pravdo et al. \cite{Pravdo}), 
the duration of the microlensing perturbations is
just between a few hours and a few days.

Since ground-based monopupil telescopes were thought to be limited
by atmospheric noise and therefore have rather poor precision (Lindegren
\cite{Lindegren}), 
only the use of high-precision optical interferometers such
as SIM (Boden et al. \cite{Boden}; Paczy\'{n}ski \cite{Paczynski}), 
Palomar Testbed Interferometer (Lane \& Muterspagh \cite{Lane}),
Keck (Colavita et al.
\cite{Colavita}), and VLTI (Wittkowski et al. \cite{Wittkowski}) was
considered so far for carrying out observations with such precision.
A quick progress in adaptive optics, and its application to astrometry
(Pravdo et al. \cite{Pravdo06}) leads to impressive results. 
However, the theoretical considerations by Lazorenko \& Lazorenko 
(\cite{Lazorenko4})
and a first set of corresponding data 
(Lazorenko \cite{Lazorenko6}, hereafter Paper I)
demonstrated that astrometric measurements with very large ground-based
telescopes are {\em not} atmospheric-limited. Instead, a high
precision beyond this limit is achievable by elimination of atmospheric
image motion at the data-reduction phase. For that purpose, 
a field of reference stars is configured into a virtual
high-pass filter that  absorbs most significant low-frequency
modes of the image motion spectrum.
For telescopes with a fully illuminated entrance pupil, 
the atmospheric error $\sigma_{\rm{at}}$
rapidly decreases with the increase of the telescope entrance pupil
$D$. At $D \to \infty$, the asymptotic dependence is
\begin{equation}
\label{eq:int}
\sigma_{\rm{at}} =B_k R^{11/6} D^{-3/2} T^{-1/2}
\end{equation}
where $T$ is the integration time,  $R$ is the angular reference field
radius, and $B_k$ is the filter parameter. 
This relation involves an additional factor $\sqrt{R/D}$ as compared
to that predicted
by Lindegren (\cite{Lindegren}),
so that large telescopes with $D=8-10$~m and above are
favourable. A specific amplitude apodization
to the entrance pupil of future telescopes may further suppress the atmospheric
error to $\sigma_{\rm{at}} \sim  R^{k/2} D^{-k/2+1/3} T^{-1/2}$ where
$k$ is the reduction parameter.

The astrometric performance of the VLT on very short time-scales could have
been estimated from test reductions of a four-hour series of FORS2 frames obtained 
in the region of the Galactic bulge (Paper I). 
A positional accuracy of 300 $\mu$as per single $R$=16 mag star measurement
at 17 sec exposure was shown to be limited only by the number of photons received 
from the observed star. 

The current study extends the certification of high-accuracy astrometric performance 
with the VLT to time scales of 1--4 days,
which is sufficient for monitoring astrometric planetary deviations
in microlensing events. In Sect.~2 of this paper, we lay out the
observational basis of this investigation, describe the image processing, and
show how the stellar centroids are computed. Particular
problems in astrometric reductions that are based on background reference
stars with unstable positions, and a new reduction method are
discussed in the subsequent Sect.~3. 
Reduction of the observational data with this method is described in Sect.~4.
Sect.~5 focusses on
the noise and correlations in the measurements that, if exist, can be 
related to the different parts of the telescope (optics, mechanics, electronics), 
to the atmosphere, etc., preventing statistical 
improvement of the accuracy by means of simple accumulation of the number of frames.

\section{Observations and computation of centroids}
For our study, we were able to use VLT frames that are available
in the ESO Archive: those obtained with the FORS1 camera of UT1
in the sky region near the neutron star RX J0720.4-3125
under the ESO program 66.D-0286 (Motch et al. \cite {Motch})
were found to be best-suited for our purposes. 
Images were acquired during 4 nights from 20 to 23 December 2000 (40 frames)
and 4 nights from 29 December 2002 to 02 January 2003 (25 frames). During each
5 to 12 frames were obtained in $B$ filter with
$\sim 10$~mn exposure time and FWHM varying 
from 0.45 to 0.85$\arcsec$, where the high-resolution mode
with 0.10$\arcsec$/px scale and about $3.3 \times 3.3\arcmin$ 
field-of-view was used.   

These data enable us to carry out a detailed and comprehensive
study of the astrometric precision that is achievable with the VLT/FORS1 at
short time scales. 
Moreover, our analysis serves as a test reduction for future observations.
Unfortunately, between some frames, the images turned out to 
be displaced by more than 200~px.
For that reason, peripheral stars frequently needed to be omitted
for much jittered frames. Moreover, for astrometric purposes,
the choice of the $B$ filter is not optimal, since it results in
larger residual atmospheric chromatism as compared to $R_{\rm{special}}$. 
However, it were just these unfortunate circumstances that forced us
to look for a reduction that significantly improves the quality. 

Observations were obtained with the Longitudinal Atmospheric Dispersion 
Corrector (LADC), which is a two-prism optical device 
(Avila et al. \cite{Avila}) that allows to correct for
atmospheric differential chromatic refraction (DCR). 
DCR is caused by the atmosphere, which
acts like a prism with a certain dispersion and thereby
smears and displaces stellar images (e.g. Monet et al. \cite{Monet}).  
The size of this effect depends on the zenith distance and the colour of
the observed star and can be modelled as shown in Eq.~(\ref{eq:h}).

Raw images were debiased and
flat-fielded using master calibration frames taken from 
the ESO Archives. Positions $x$, $y$ of stellar centroids were computed using
the profile fitting technique that had already been applied earlier for the high-precision astrometric reduction of FORS2 images (Paper I). Stellar profiles in 
$10 \times 10$~px windows were fit by a sum of three modified elliptic Gaussians
involving 12 free model parameters altogether. 
The dominant Gaussian centered at $x$, $y$ with extent parameters
$\sigma_G^x$, $\sigma_G^y$ along corresponding coordinate axis and futher
specified by its orientation angle, was considered to contain 
a flux $I$ mounting to about 2/3 of
the total light received from the star.
Other auxiliary Gaussians allowed profile fitting to the photon noise limit
at which residuals of pixel counts from the model are characterized by
$\chi^2 \approx 1.0/$px  (except for central parts of images of bright stars).
Objects with $\chi^2 \ge 3.0/$px deviations were rejected and
only unsaturated images with $B>18$~mag were processed.

Due to the complex shape of the PSF, the precision $\varepsilon$ of the
image centroiding was obtained by numerical simulation, 
for which a set of randomly generated star images with profiles comparable to the observed ones was created and where Poisson noise was added to pixel counts.
This simulation shows that $\varepsilon$ can be well-approximated by
\begin{equation}
\label{eq:accur}
\varepsilon = \frac{ \theta}{2.46\sqrt{I}}(1+ \alpha_0 I^{-0.7}), 
\end{equation}
where $\theta$ denotes the seeing (which turned out to be  
$\theta=3.10\sigma_G$ in our three-component model),
$I$ refers to the electron count, and  
$\alpha_0$ is a characteristic constant that depends on the background noise.
For the model background fluxes $b_0$ 
of 810 or 540~e$^-$/px (average values at 2000
and  2002-2003 epochs), $\alpha_0$ was found to be 1150 or 887, respectively.
Eq.~(\ref{eq:accur}) becomes an exact relation only for a fixed average seeing 
$\theta_0=0.608\arcsec$ at which $\sigma_G=1.96$~px, but it provides a
reliable approximation for variable FWHM of that order.

\section {The reduction process}
\subsection {Filtration of atmospheric image motion}

Let us shortly recall the previously discussed process of attenuation
of the image motion spectrum (Lazorenko \& Lazorenko \cite {Lazorenko4}),
which is based on the virtual symmetrization of the reference field.
Consider some region of the sky with $N$ stars being imaged in $m=0,1 \ldots  M$ consecutive frames, where $x_{lm}$, $y_{lm}$ are the centroids 
of $l=1,2 \ldots N_m$ stars measured in frame $m$.
For various reasons (jittered images, cosmics, outliered data, etc.) 
a significant fraction of centroids cannot be obtained from every frame,
therefore $N_m \leq N$.
Here and below, indices $i$, $j$, $l$  refer to
the target, reference, and star of any type, respectively,
and $m=0$ is assigned to the reference frame. 
Upper indexes $x $, $y $, if used, refer to the coordinate axes.
A list of variables  most frequently used in the Paper
with a  short description is given in Table~\ref{var}.
\begin{table}[tbh]
\caption [] {A list of frequently used variables}
\begin{tabular}{@{}ll@{}r@{}}
\hline
\hline
quantity & definition  & first use,\\ 
     &                &   equation  \rule{0pt}{11pt}\\
\hline
$x$, $y$    & measured star positions        & (3) \\
$X$, $Y$   & positions $x$, $y$ filtered of image motion  and  given & (3) \\
&         in the system of local reference stars with &\\
&  ''fixed'' positions&\\
$\lambda ^x$, $\lambda ^y$ &  same as $X$, $Y$ but given in a single all-frame&\\
&             system with ''floating''  reference stars & (9)\\
$\xi^x$, $\xi^y$  & displacement (average for a given frame series) &(14)\\
&         of the star from its position in reference frame, &\\
&     for astrometric reasons  (proper motion etc.)&\\
$\hat{\xi}^x$, $\hat{\xi}^y$  & same as $\xi^x$, $\xi^y$ but given  with
          respect to  the & (10)\\
&         mathematical expectation of the star position in &\\
&         reference frame & \\
$\rho$ &  chromatic coefficient;  atmospheric displacement  & (12)\\
&  of the star image is proportional to  $\rho$  &\\
$d$     & chromatic coefficient; describes compensating & (12)\\
&            action of the LADC  & \\
$h$  & differential chromatic displacement (DCR effect) & (12)\\
&            of the star image  & \\
${\mathcal M}$  & atmospheric image motion effect in  $x$, $y$  & (10)\\
$\Delta{\mathcal M}$ & residual atmospheric image motion in $X$, $Y$ & (14)\\
$e$   & error of photocenter measurement&      (10) \\
$\varepsilon $ & variance of $e$   &  (1,25) \\
$u$    &  residuals of conditional equations (23) & (26)\\
${\mathcal N}$ & noise in observations that enters Eqs.(23) &
              (17)\\
$\sigma^2$  &  observed variance of ${\mathcal N}$ found from
          residuals $u$       & (26)  \\
\hline
\end{tabular}
\label{var}
\end{table}

The differential position $X_{im}$ of the $i$-th target star in the $m$-th frame
is obtained from CCD coordinate differences 
$x_{im}-x_{jm}$ of the target and field star centroids. More specifically,
it is given as the weighted average  
\begin {equation}                               
\label{eq:xdef}
X_{im}=  {\sum \limits _{j=1}}'  a_{ij}(x_{im}-x_{jm})=
       x_{im} - {\sum \limits _{j=1}}' a_{ij}x_{jm}, \; m=1,2 \ldots M
\end{equation}
of these differences, where the weights $a_{ij} $ satisfy the
normalization condition
\begin{equation}                               
\label{eq:norm}
 {\sum \limits _{j=1}}'  a_{ij}=1
\end{equation}
and are chosen, so that the effect of image motion effect in $X_{im}$ is minimized.
The prime indicates that the summation is carried out over the subset 
$\Omega_{im}$ of stars used as reference for $i$-th star only, where the
index $j=i$ is omitted. The corresponding positional $y$-coordinate $Y_{im}$ is
obtained in analogy to Eq.~(\ref{eq:xdef}). 

Atmospheric image motion displaces each position $x _{im}$  by a small
angle ${\mathcal M}_{im}$, which results in $X_{im}$  being displaced
by $\Delta{\mathcal M}_{im}={\sum  _{j}} ' a_{ij}
({\mathcal M}_{im}-{\mathcal M}_{jm}) $.
The spectral power density $G (q)$ of $\Delta{\mathcal M}_{im}$ 
in the space of spatial frequencies $q$ is the product of two factors.
The first factor $F'(q)$ depends on  $D$,  exposure time $T $, and 
properties of atmospheric turbulent layers generating image motion.
The second factor depends only on the geometry
of background star positions relative to the target $i$, and
can be expanded in a series of even powers of $q$. Thus,
\begin{equation}                               
\label{eq:g}
G(q)=  F'(q) \sum_{s=1}^{\infty} q^{2s} F_{2s}(\tilde{x}_i,\tilde{y}_i;
                                  \tilde{x}_1,\tilde{y}_1
		\ldots \tilde{x}_N,\tilde{y}_N)\,,
\end{equation}
where  $ \tilde {x} $, $ \tilde {y} $ are
positions not affected by image motion (at zero turbulence).
The coefficients $F_{2s} $ are quadratic functions
of coordinate differences that fastly decrease with $s$ and take the form
\begin{equation}
\label{eq:f}
F_{2s}= \sum  \limits_{ \begin{array}{c}
		^{\alpha,\beta =0,}
		_{\alpha+\beta = s} 
		  \end{array} }   ^{s}
          A_{\alpha \beta}^{(2s)}
	\left [ {\sum \limits _{j}} ' a_{ij}
	(\tilde{x}_i-\tilde{x}_j)^{\alpha}(\tilde{y}_i-\tilde{y}_j)^{\beta }
       \right ] ^2
\end{equation}
where $ A _ {\alpha \beta} ^ {(2s)} $ are the characteristic constants.
With the choice of 
coefficients $a _{ij} $, all sums in brackets can be turned to zero
for a specific $ A _ {\alpha \beta} ^ {(2s)} $.
In this way, the leading term $F_2 $ and, possibly, several subsequent $F _ {2s} $ terms are eliminated, so that $G(k)$ will depend largely on the first non-zero component $F _ {k} $ of some optional  mode $k=4, 6 \ldots  $.
For this purpose, the coefficients $a _ {ij} $ should satisfy
each of $k (k+2) /8-1 $ conditions
\begin{equation}
\label{eq:systa1}
\begin{array}{lr}
{\sum \limits _{j}} ' a_{ij} (x_{i0}-x_{j0})^{\alpha} 
         (y_{i0}-y_{j0})^{\beta}=0,& \; 
		\alpha + \beta = 1 \ldots  \frac{k}{2}-1,   \\
\end{array}
\end{equation}
where $k \geq 4 $, $ \alpha $ and $ \beta $ are non-negative integers, and 
positions $ \tilde {x} $, $ \tilde {y} $ 
are substituted by respective
measured positions of stars in the reference frame.
The coefficients $a _ {ij} $ are determined 
for each target $i$ by solving  Eqs.~(\ref {eq:systa1}) with normalizing
condition (\ref {eq:norm}), provided that at least
$N ' =k (k+2) /8 $ reference stars are available. Because usually $N_m \gg 
N ' $, the system of equations (\ref {eq:systa1}) 
is redundant and  solvable with some useful
restrictions on $a _ {ij} $, for example Eq.~(\ref {eq:min}).
Due to the elimination of principal modes of the image motion spectrum, 
the resulting values of $X _ {im} $ are characterized by 
the residual image motion $\Delta{\mathcal M}_{im}$ that involves only
uncompensated spectral modes of orders $s \geq k/2 $ and therefore
is much smaller than its initial value. 
Approximately, the variance of  residual image motion is expressed by
Eq.~(\ref {eq:int}),
where $R$ is to be substituted by  the radius $R_i $ of local reference field
for $i$-th target. The advantage of using larger orders in $k$ is reflected in
the coefficient $B_k$ which decreases with $k$. In particular,
for $k=12$, its value is only 20\% of that measured at the 
VLT with $k=4$ (Paper I).

Next, let us introduce polynomial base functions  
$ f_{1l}=1$,   $ f_{2l}=x_{l0}$, 
$ f_{3l}=y_{l0}$, $\ldots$ $ f_{wl}=x_{l0}^{\alpha}y_{l0}^{\beta}$
of coordinates $x _ {l0} $, $y _ {l0} $ referring to the  $l$-th  star in 
the reference frame, with  indices $w=1,2 \ldots N ' $  equal
to the  sequential number of combinations of $ \alpha $ and $ \beta $ in equations (\ref {eq:systa1}). 
Taking advantage of the equality
$ {\sum  _{j}}  ' a_{ij} (x_{i0}-x_{j0})^{\alpha}(y_{i0}-y_{j0})^{\beta}
= x_{i0}^{\alpha}y_{i0}^{\beta} - 
{\sum  _{j}}  ' a_{ij} x_{j0}^{\alpha}y_{j0}^{\beta}$,
we can rewrite equations (\ref {eq:norm})  
and (\ref {eq:systa1}) in  terms of functions $f _ {wl} $ as
\begin{equation}
\label{eq:systa2}
\begin{array}{lr}
 {\sum \limits _{j}} ' a_{ij} f_{1j} =1, \; &
 {\sum \limits _{j}} ' a_{ij} f_{wj} =f_{wi}, \;
                         w=2,3 \ldots N'
\end{array}
\end{equation}

\subsection{Conversion from $X$, $Y$ to displacements $\lambda ^x$, 
$\lambda ^y$} 
The image displacement $X _ {im}$ of the $i$-th star between
frame $m$ and the reference frame, filtered from low-order image motion and
systematic distortions of the field, is measured
relative to a set $\Omega_{im} $ of background stars 
whose positions are assumed to be fixed. However, since such are not available,
we bypass this limitation by introduction of astrometric displacements
${\lambda}$ that are related  
to all stars in the frame without distinction between
reference and target objects. 
We consider $X _ {im}$ and ${\lambda}$ related by
\begin{equation}
\label{eq:x}
x_{lm}= x_{l0}  +P_{lm}+ {\lambda}_{lm},
\end{equation}
associating the measured position of stars in frame $m $ with their
position in the reference frame.
Here, $P _ {lm} = \sum_w c _ {wm} f _ {wl} $ is a polynomial 
function that contains powers of $x $, $y $ not higher than $k/2-1 $
and measured at discrete set of points $x_{l0}$, $y_{l0}$. 
Via coefficients $c_{wm} $,
this function  describes the systematic variation of  stellar positions
with the change of optical aberration and other similar effects.
The second quantity
\begin{equation}
\label{eq:lmu}
{\lambda}_{lm}= \hat{\xi}_l+e_{lm} - e_{l0} + h_{lm}-h_{l0} 
+{\mathcal M}_{lm}- {\mathcal M}_{l0} 
\end{equation}
is orthogonal to  $P_{lm}$ and satisfies conditions
\begin{equation}
\label{eq:m3}
\begin{array}{lrr}
  \sum_i \lambda_{im}^{x} f_{wi} =0, & \sum_i \lambda_{im}^{y} f_{wi} =0, &
  \; w=1,2 \ldots N'
\end{array}
\end{equation}
In fact, Eqs.~(\ref {eq:x}) and (\ref {eq:m3}) give an expansion of
$x_{lm}$  into power series of $ f_{wl}$, with $\lambda_{lm}$
representing the remainders. $\lambda_{lm}$ includes
the ''science'' astrometric signal $\hat{\xi}_l$ (sum of 
proper motion, parallax, probable microlensing and
reflex motion caused by the planetary companion)
measured relative to the reference frame,
random errors of  photocenter measurement
$e _ {lm} $,
the atmospheric chromatic displacement $h _ {lm} $ corrected by LADC, and
the atmospheric image motion ${\mathcal M}_{lm}$. 
These quantities (or their combinations) are also orthogonal to $ f_{wl}$,
with their low-order $w \leq N'$ 
expansion terms included in $P_{lm}$ via coefficients $c_{wm}$.

The components of $h _ {lm} $ along the $x $- (opposite to right ascension) 
and $y $- (along declination) axes 
are given by (Paper I) 
\begin{equation}
\label{eq:h}
\begin{array}{lr}
 h_{lm}^x= \rho_{l}\tan{z_m} \sin{\gamma_m} + 
 d_{l} \tan{\zeta_m} \sin{\gamma_m} & {\rm {for}} \; x-{\rm {axis}} \\

 h_{lm}^y= -\rho_{l}\tan{z_m} \cos{\gamma_m} -
 d_{l} \tan{\zeta_m} \cos{\gamma_m} & {\rm {for}} \; y-{\rm {axis}}
\end{array} 
\end{equation}
Here, $ \rho _ {l} $ is the coefficient of differential chromatic
displacement for the $l $-th star, $d _ {l} $ is 
a negative displacement produced by LADC
to compensate for DCR effect, approximately of the same absolute size but
opposite sign, $z_m $ is a zenith distance
at the midpoint of exposure of the $m$-th frame, 
$ \gamma_m $ is the parallactic angle,
and $ \zeta_m $ is the zenith angle of the LADC setting, which remains 
the same for a single series of observations.

Substitution of Eqs.~(\ref {eq:x}) for $x _ {im} $ and $x _ {jm} $ 
into Eq.~(\ref {eq:xdef}) leads to elimination of the
polynomials $P_{lm}$, while fulfilling the conditions given by
Eq.~(\ref {eq:systa2}). Taking into account that
$X_{i0}={\sum  _{j}} ' a_{ij} (x_{i0}-x_{j0})=0$ due to Eqs.~(\ref {eq:systa1}),
we obtain
\begin{equation}
\label{eq:lambda1}
 \lambda _{im}^x - {\sum \limits _{j}}'a_{ij} \lambda_{jm}^x=X_{im}\,.
\end{equation}
An analogous expression for transformation of $Y_{im} $ to $\lambda_{im}^y$
results for measurements along the $y$-axis. 
Eqs.~(\ref {eq:lambda1}), for each frame $m $, form
a system of $ i=1 \ldots N_m $  linear equations with $N_m $ 
unknowns $ \lambda _ {lm}^x $.
The solution for a given frame $m $ is found irrespective of solutions 
for other frames.
The matrix of the system (\ref {eq:lambda1}), however, turns out to be degenerated
due to $ N' $ linear dependences (\ref {eq:systa1})   
between its columns (coefficients $a _{ij} $),
and consequently it can be solved only using the same number of 
auxiliary restrictions, for example (\ref {eq:m3}).   

Below, we discuss two approaches to the extraction of 
$ \hat{\xi}^x_i$, $ \hat{\xi}^y_i$, $\rho_i$, and $d_i$ from $\lambda_{im}$.

\subsection{A case of fixed local reference star positions}
In the case of small background star displacements $\lambda_{jm} $,
and due to the averaging effect at summation,  a term 
${ \sum  _{j}}'a_{ij} {\lambda}_{jm}$ in Eq.~(\ref{eq:lambda1}) can be 
discarded
as negligibly small. In fact, this assumes a fixed  position
of local reference stars $\lambda_{jm}=0 $ 
that leads to the most simple result $\lambda_{im}^x= X_{im}$. 
For each target $i$, we are facing a system  of $2M $ equations
\begin{equation}
\label{eq:stable}
\begin{array}{l}
 {\xi}_{i}^x + h_{im}^x - h_{i0}^x 
=
 X_{im} - e_{im}^x - \Delta{\mathcal M}_{im}^x  \\
 {\xi}_{i}^y + h_{im}^y - h_{i0}^y 
=
 Y_{im} - e_{im}^y - \Delta{\mathcal M}_{im}^y  \\
 m=1,2 \ldots M, 
\end{array} 
\end{equation}
for determining
${\xi}_{i}^x= \hat{\xi}_{i}^x-e_{i0}^x- \Delta{\mathcal M}_{i0}^x$, 
$ {\xi}_{i}^y =  \hat{\xi}_{i}^y-e_{i0}^y- 
\Delta{\mathcal M}_{i0}^y$, $\rho_i$, and $d_i$. 
This system can be solved by the least-squares method for each
target object independently, 
with reference to its own subset $\Omega_{im}$ of reference stars, 
considering $e _ {im} $ and $ \Delta {\mathcal M} _ {im} $
as random (with respect to index $m $) 
errors with variances
$ \varepsilon _ {im} ^2 $ and $ \sigma^2 _ {\rm {at}} $, respectively.
Errors $e _ {jm} $ of reference star photocentre determinations
add a further uncertainty to the phase of compution of
$X _ {im}$ by means of Eq.~(\ref {eq:xdef}). 
A variance of this error
$\sigma^2_{\rm {rf}}={\sum  _{j}} ' a_{ij}^2 \varepsilon_{jm}^2$,
equal to the variance  
of a term  $ {\sum  _{j}} ' a_{ij} x_{jm}$ in Eq.~(\ref{eq:xdef}),
can be reduced applying restrictions
\begin{equation}
\label{eq:min}
 {\sum \limits _{j}} ' a_{ij}^2 \varepsilon_{jm}^2 = \mbox {min}  
\end{equation}
on solutions $a _ {ij} $ of the system (\ref {eq:systa1}).

As it was mentioned above, 
geometric field distortions (their variations in time), 
up to order $k $,
are excluded along with image motion.

This simple method was applied
for the test reduction of four-hour series of frames obtained at 
VLT/FORS2 during one night (Paper I). 
By filtration of atmospheric image motion 
carried out with  $k $ up to 12, 
300~$\mu$as precision of a single position for  bright stars was achieved,
which is near to the accuracy of photocentre measurements.
Approximation $ \lambda _ {jm} =0 $ was valid due to a small 
residual DCR effect for frames obtained with the
$R _ {\rm {special}} $ filter.

\begin{figure}[htb]
\begin{tabular}{@{}c@{}c@{}}
{\includegraphics*[bb = 53 49 225 209, width=4.6cm,height=4.9cm]{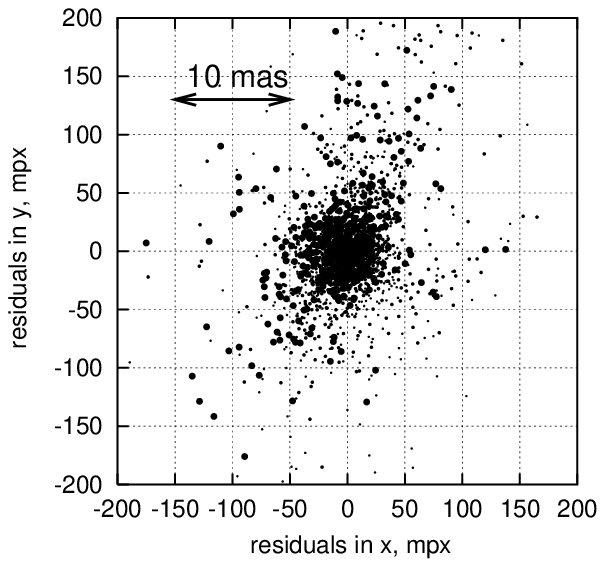}}&
{\includegraphics*[bb = 58 49 225 209, width=4.2cm,height=4.9cm]{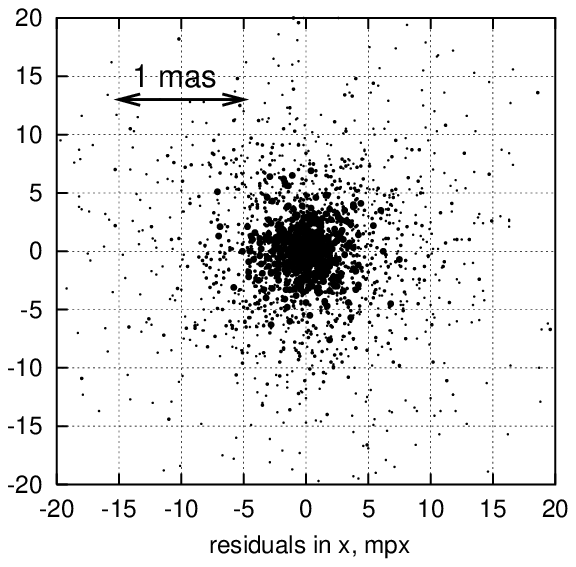}} \\
\end{tabular}
\caption {  $x$, $y$ residuals of Eqs.~(\ref{eq:stable})
for 25 images computed with the assumption of ''fixed''  background star positions  
({\it left panel}) in comparison to the exact solution ({\it right})
for epochs of reference and processed frames being separated by two years.
Dot sizes are proportional to the stellar magnitude (only bright $B<$21.5~mag stars
are shown). 
Note the difference in scales (1~mpx=100~$\mu$as).}
\label{xy}
\end{figure}

%%%%%%%%%%%%%%%%%%%%%%%%%%%%%%

\subsection{Reduction for floating reference stars}

For the processing of sky images in the $B$-filter, the use of the technique descrived
about is however limited, even for frames taken within the same night. In this case, 
an excess of positional errors over $\varepsilon_i$ by a factor 3--5 results,
because the underlying assumption of $\lambda_{jm}=0$ does not hold
due to large ($\sim$2--10~mas/hour) chromatic displacement of images.
For frames taken two years after the reference frame, which results in
significant additional proper motion displacements, 
the  solution of Eqs. (\ref{eq:stable})
produces unacceptably large residuals (Fig.\ref{xy}a).

The exact solution of Eqs.~(\ref {eq:m3}--\ref {eq:lambda1}) 
ensures a linearity and
uniqueness of the transformation from $X _ {im} $  to $ \lambda _ {im} $ 
even for moving ($ \lambda _{im}\neq 0 $) reference stars.
Therefore, although $X _{im} $ are determined relative to local reference fields 
$\Omega_{im}$
centered at the $i$-th target and depend on the size $R_i$ of these
fields, the final solutions 
$ \lambda _ {im} $  are consistent within the global 
reference  field of size $R$ and do not depend on the choice of 
initial $R_i $ values.
Since usually $R > R_i $, the transformation from $ X_{im} $ to $ \lambda _ {im} $ 
increases the atmospheric image motion component
in $ \lambda _ {im} $,  which should be compensated for
by moving to larger $k$.

Note that  due to conditions (\ref {eq:systa2}),
both $ \lambda_{im}$ and  $ \lambda_{im}+\varphi_{im} $
are the solutions of a system (\ref {eq:lambda1})
if $ \varphi _ {im} $  are functions 
\begin{equation}
\label{eq:fi}
\begin{array}{lrr}
 \varphi_{im}^x= \sum \limits _{w=1}^{N'}E_{wm}^x f_{wi}, &&
\varphi_{im}^y= \sum \limits _{w=1}^{N'}E_{wm}^y f_{wi} \\
\end{array}
\end{equation}
with arbitrary coefficients $E _ {wm} ^x $ and $E _ {wm} ^y $. Therefore,
for splitting $ \lambda _ {im} $ into components we can use
\begin{equation}
\label{eq:lambda2}
\begin{array}{l}
 {\xi}_{i}^x + h_{im}^x - h_{i0}^x 
 + \varphi_{im}^x=
 \lambda_{im}^x    + {\mathcal N}_{im}^x \\

 { \xi}_{i}^y + h_{im}^y - h_{i0}^y 
 + \varphi_{im}^y=
 \lambda_{im}^y +{\mathcal N}_{im}^y  \\
 m=1,2 \ldots M, \; \; i=1,2 \ldots N
\end{array} 
\end{equation}
where ${\mathcal N}_{im}= -e_{im}-\Delta{\mathcal M}_{im}$ is a noise
with variance $ \varepsilon _{im} ^ {2} + \sigma _{\rm {at}} ^2 $.
Eqs.~(\ref {eq:lambda2}) can then be solved as a system
of $2\sum N_m $  conditional equations with $4N+2M N'$  unknowns 
${ \xi}_{i}^x$, 
${ \xi}_{i}^y$, $\rho_i$, $d_i$, $E_{wm}^x$, and  $E_{wm}^y$.

Because for most not very faint 
targets, $ \sigma _ {\rm {at}} \ll \varepsilon _ {im} $,
the least-squares solution is found with weights
$g_{im} = \varepsilon_{im}^{-2}$. 
Also, due to the dominant contribution of bright stars,
we can safely assume $g_{im}^{2} \approx I_i/\theta_{im}^2$ which follows from the
approximation given by  Eq.~(\ref{eq:accur}) for these stars.

Linear relations between some parameters of a system
(\ref {eq:lambda2}) do not permit to separate them completely. 
For example, systematic variations
of $ {\xi} _ {i} ^x $, $ \rho_i $, and $d_i $ quantities 
over a field imitate the behaviour
of $ \varphi _ {im} ^x $ functions. Associations of this type are  eliminated
by applying $N ' $ orthogonalizing restrictions
\begin{equation}
\label{eq:restrictions}
\begin{array}{lll}
 \sum \limits _i { \xi}_{i}^x f_{wi}I_i =0,       &
  \sum \limits _i{ \xi}_{i}^y f_{wi}I_i =0,  & \\

  \sum \limits _i \rho_{i}f_{wi}I_i =0, &
  \sum \limits _i d_{i}   f_{wi}I_i =0,   &
  w=1,2 \ldots N'                  \\

\end{array} 
\end{equation}
The restrictions (\ref {eq:restrictions}) permit us to recover only those 
components of $ \rho_i $, $d_i $, $ {\xi} _ {i} ^x $, and 
$ {\xi} _ {i} ^y $, that do not correlate with 
functions $ f _ {wi} $.

An essential drawback of a straightforward solution of  
Eqs.~( \ref {eq:fi} -\ref {eq:restrictions}) is a large number of 
unknowns $E _ {wm} ^x $ and $E _ {wm} ^y $, that cause a certain
instability of a solution.
For example, the number of unknowns for Dec 2000 frames is 
$2MN'=2 \cdot 40 \cdot 36 \approx 2900$ at $k=16$,
which is comparable to the number of measurements $2 \sum N_m \approx 9100$.

However, there is a simpler solution.
Consider equation 
with a diagonal element  $E _ {wm} ^x $
of the normal system of Eqs.~(\ref {eq:lambda2}):
\begin{equation}
\label{eq:no}
\begin{array}{l}
\sum \limits _{i \in \omega_m}({\xi}_{i}^x + h_{im}^x - h_{i0}^x
     )f_{wi}I_i + 
          
     \sum \limits _{i \in \omega_m}f_{wi}I_i \sum \limits _{\bar{w}}
            E_{\bar{w}m}^x f_{\bar{w}i}\\
            =    \sum \limits _{i \in \omega_m} \lambda_{im}^xf_{wi}I_i
\end{array} 
\end{equation}
where summation is performed only over a limited sample $ \omega_{m}$ of stars
measured
in  the $m $-th  frame and, since FWHM$\approx$const within the frame, the
approximation $g_{im}=I_i$ was used. 
Adding the sum
\begin{equation}
\label{eq:L}
\Lambda _{wm}^x= \sum \limits _{i \in \mbox{ \small{not} } 
     \omega_m}({\xi}_{i}^x + h_{im}^x - h_{i0}^x
     )f_{wi}I_i 
\end{equation}
formed over stars {\it not} measured in the frame $m $ to both sides of Eq.~(\ref {eq:no}), and taking into account
(\ref{eq:restrictions}), we note that the first component in 
Eq.~(\ref{eq:no}) turns to zero. Therefore
\begin{equation}
\label{eq:L2}
\sum \limits _{i \in \omega_m}f_{wi}I_i \sum \limits _{\bar{w}}
            E_{\bar{w}m}^x f_{\bar{w}i}=
            \sum \limits _{i \in \omega_m} \lambda_{im}^xf_{wi}I_i
       +\Lambda _{wm}^x\,.
\end{equation}
If a solution of the system (\ref {eq:lambda1}) for the $m $-th frame 
is found that fulfills the conditions 
\begin{equation}
\label{eq:L3}
 \sum \limits _{i \in \omega_m} \lambda_{im}^xf_{wi}I_i=
       -\Lambda _{wm}^x  \; \; w=1,2 \ldots N',
\end{equation}
instead of those of Eq.~(\ref{eq:m3}), the right part of the system given by
Eq.~(\ref {eq:L2}) turns to zero, thus yielding a unique solution
$E _{wm} =0 $. 
Similar considerations for the $y$-component of the data yield $E _{wm} =0 $, 
hence $ \varphi _ {im}^x = \varphi _ {im}^y =0 $. 
The elimination of the large number ($2MN ' $) of unknowns $E _ {wm} ^x $ and
$E_{wm}^x$ in Eq.~(\ref{eq:lambda2}) occurs due to imposing an equivalent
number of restrictions (\ref{eq:L3}), thereby increasing the
effective degree of freedom of the system given by Eq.~(\ref{eq:lambda2}) 
to $2\sum N_m -4N$.
Furthermore, this splits the system given by Eq.~(\ref{eq:lambda2}) into $N $ 
independent subsystems of $ 2 n_i $ equations ($n_i $ is 
number of observations of the $i$-th star)
\begin{equation}
\label{eq:work}
\begin{array}{l}
 {\xi}_{i}^x + h_{im}^x - h_{i0}^x 
 =
 \lambda_{im}^x + {\mathcal N}_{im}^x   \\

 { \xi}_{i}^y + h_{im}^y - h_{i0}^y 
 =
 \lambda_{im}^y + {\mathcal N}_{im}^y  \\
\end{array} 
\end{equation}
which is easily solved for ${ \xi}_{i}^x$, 
${ \xi}_{i}^y$, $\rho_i$, and $d_i$ for each of the star observed independently.
With this processing technique,  $ \Lambda _ {wm} $ 
entering Eq.~(\ref {eq:L3}) are calculated by iteratively refining 
$ {\xi} _ {i} ^x $, $ {\xi} _ {i} ^y $, $ \rho_i $, and $d_i $. 
Iterations are however not required if 
each star is measured in all of the frames since in this case
the identity $ \Lambda _ {wm} ^x =\Lambda _ {wm} ^y=0 $ holds due to
Eq.~(\ref{eq:restrictions}).

Some comments should be made on the nature of 
${ \xi}$ values computed not in the absolute reference
frame, but only relative to a limited set of $N$ nearby stars. 
Systematic differences between
${ \xi}$ and values ${ \xi}_{\rm{abs}}$ referring to the absolute or 
global frame, are given by
\begin{equation}
\label{eq:abs}
 \xi _{{\rm {abs}},i}= { \xi}_{i}+ \sum_w G_w f_{wi}
\end{equation}
which incorporates a polynomial function $\sum_w G_w f_{wi}$
of $N'$ order with coefficients $G_w$. 
The values $G_w$ should confirm the restrictions given by Eq.~(\ref{eq:restrictions})
which now take a form 
$\sum_i({ \xi}_{{\rm{abs}},i} - \sum_w G_w f_{wi})^2 I_i= \mbox{min}$,
or $\sum_i { \xi}_{i} ^2 I_i= \mbox{min}$. Eq.~(\ref{eq:abs}) thus is
an expansion of ${ \xi}_{{\rm {abs}},i} $ into base functions $f_{wi}$,
and $ \xi_{i} $ are the residuals of the expansion.
The increase of $N'$ (or $k$)
leads to the increase of polynomial modes excluded from $\xi_{\rm{abs}}$
and thus to the loss of information retained in $\xi$. Reduction
therefore should use the lowest $k$ that still ensures acceptably
small $\sigma_{\rm {at}}$. Above considerations, of course,
are fully applicable to $ \rho_i $ and $d_i $ parameters as to the
remainders of corresponding polynomial expansions.

\section{Reduction of measured images}

An image obtained on 21 Dec 2000 at an hour angle near to
the mean hour angle of other images was taken as a reference frame.
Only stars in the central region of the sky with an
$R=1000 $~px radius  containing $N=169 $ 
star images down to $B=24$~mag were selected for the reduction. 
Peripheral stars which due to
a strong jittering were often beyond frame boundaries thus were rejected.
Nevertheless, the number of stars 
measured varied among the frames and reached down to $N_m \approx 90 $.

We have taken advantage of the facts that the colours of the observed stars
do not vary significantly on short time-scales (few days) and 
positions of most stars are only slightly affected by proper motions and
parallaxes. Therefore frames were combined into series
within annual epochs, which statistically improved the accuracy of model 
parameters determination. We therefore organized our data into
two series of 39 and 25 frames for 2000 and
2002-2003 epochs, respectively, each one representing four nights.
Astrometric reduction was carried out with different choices of the
parameter $k $, where the accuracy of results became sufficiently good at $k=16$,
which was adopted for the final reduction.
Our iterative procedure involved the computation of $ {\xi} _ {i} ^x$, 
${ \xi}_{i}^y$, $\rho_i$, 
as well as $d_i $ and the refinement of $\Lambda_{im}$ quantities
calculated on the basis of the former parameters.
The number of outliers at each coordinate was limited to
1.25\% of the total number of measured star images.

\begin{figure}[htb]
\includegraphics*[bb = 82 45 249 153, width=8.7cm,height=5.0cm]{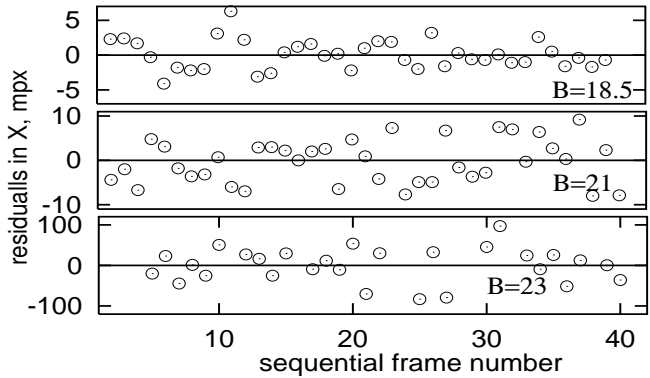} \\

\caption {  Example of ${u}_{im}^x$ residuals           
for 2000 epoch frames. }
\label{example}
\end{figure}

The residuals $u _ {im} ^x $ and $u _ {im} ^y $ 
of the conditional equations Eq.~(\ref {eq:work}) on $x $- and $y $- coordinates
are uncorrelated (Fig.\ref{xy}b) and,
besides,  substantially smaller than those obtained with the
approximation of ''fixed'' background star positions
(Fig.\ref {xy}a).
We also did not find any correlation between $u_{im}$ and the stellar position within the frame 
or the colour-dependent parameters $\rho_i$ and $d_i$. 
As a simple illustration for a further discussion,  Fig.\ref{example}
presents examples of $u _ {im} ^x$ residuals for stars of different
magnitude, from bright $B=18.5$ to faint $B=23$~mag. A difference in the
data point scattering reflects the dependence of photocenter error on the
light signal. No apparent correlation in time is seen. More extensive
statistical analysis of $u _ {im}$ values is performed in  Sect.5.

%%%%%%%%%%%%%%%%%%%%%%%%%%%%%%

\subsection{Performance of the LADC }
\begin{table}[tbh]
\caption [] {Relative corrections to $\tan{\zeta_m}$: producing best
accuracy of observations ($C_m$)
and corresponding to a 10~mn delay in the LADC setting ($C'_m$)}
\begin{tabular}{rrr|rrr}
\hline
\hline
night & $C_m$ & $C'_m$ & night  & $C_m$  & $C'_m$  \rule{0pt}{11pt}\\ 
\hline
2000 &   &  & 2002-2003  &     &    \\
20 Dec &  -0.17 & -0.15    &  29 Dec    &  -0.06  & -0.11 \\
21 Dec &    0 & -0.02    &  30 Dec    &  0.05  & 0.09 \\
22 Dec &  0.26 & 0.20    &  31 Dec    &  -0.11  & -0.15 \\
23 Dec &  -0.11 & -0.14    &  02 Jan    &  0.02  & -0.13 \\
\hline
\end{tabular}
\label{ladc}
\end{table}

Zenith distances $\zeta_m$ of the LADC setting used in Eqs.~(\ref{eq:h})
are not specified in the fits file headers. According to the technical
documentation of the VLT,  $\zeta_m$ is
equal to the telescope zenith distance
at the midpoint exposure time of the first science image of the series and
is constant for all frames within the series. 
We however tried to 
use corrected values $(1+C_m)\tan{\zeta_m}$ instead of $\tan{\zeta_m}$
in order to improve internal precision. Coefficients $C_m$ (Table \ref{ladc})
that yield the best-achievable precision have been determined with an
accuracy of about $\pm0.03$ 
relative to some zero-point night, chosen as 21 Dec 2000.
Closer examination revealed that 
$(1+C_m)\tan{\zeta_m}$  is approximately equal to 
$\tan{\zeta'_m}$ computed with  the telescope zenith distance ${\zeta'_m}$ 
10~mn prior to the first exposure. Writing
${\zeta'_m}$ in the form $\tan{\zeta'_m}=(1+C'_m)\tan{\zeta_m}$ allowed us
to introduce  new values $C'_m$ and compute them in analogy to $C_m$.
A good consistency of $C_m$ and $C'_m$ values suggests that actual angles
${\zeta_m}$ are really close to ${\zeta'_m}$.  According to Jehin E. (private
communication, 2006),
this may occur when the LADC is not reset after finishing
preceding observations of the same target. The final reduction was 
carried out with $(1+C_m)\tan{\zeta_m}$ rather than $\tan{\zeta_m}$.

\begin{figure}[htb]
{\includegraphics*[bb = 52 49 267 194, width=8.7cm,height=6.3cm]{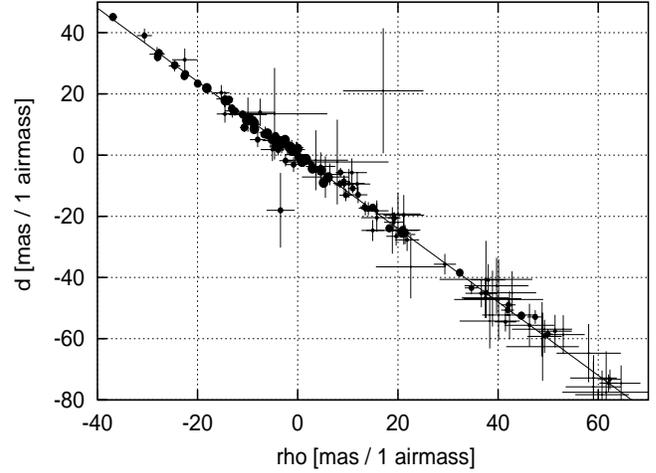}}
\caption { The LADC compensating displacement $d_i$  and
atmospheric differential chromatic refraction $\rho_i$ 
for stars of $B=18-24$~mag  (dots) with a linear (solid line)  fit.
The dot size is proportional to the stellar magnitude.
Individual error bars are shown.}   
\label{rhod}
\end{figure}
%%%%%%%%%%%%%%%%%%%%%%%%%%%%%%

An excellent performance of the LADC is demonstrated by Fig.\ref{rhod}
which reproduces the distribution of $\rho_i$ and $d_i$ (averaged over both
epochs) with error bars based on the effective precision of a single
measurement (\ref{eq:sigma}). Blue stars are shown in the upper left corner.
The distribution of stars is strongly concentrated along a line $\rho _i=-1.2 d_i$
with a  scatter that only slightly exceeds random errors.

\subsection{Effective precision of a single measurement}
Astrometric positions are affected by various sources of error.
The most important are fundamental (photon noise in the image), of
instrumental origin (optical field distortions, errors related to CCD, etc.),
induced by the atmosphere (DCR effect, image motion), or related to
inefficient reduction techniques that should take into account and exclude largest
errors. Without analyzing each source separately, some of which are rather
efficiently filtered out at the phase of the reduction, we consider
their total impact on the key system of Eqs.~(\ref{eq:work}), whose solution
yields final astrometric parameters $\xi_x$ and $\xi_y$. 
This total error in  Eqs.~(\ref{eq:work}) was denoted
as ${\mathcal N}_{im}$ and includes $e_{im}$, $\Delta{\mathcal M}_{im}$,
and a sum of other residual errors not filtered by the reduction model.
${\mathcal N}_{im}$ includes both a  random noise
and components correlated in
time. The presence of correlated components
leads to strongly biased results if the correlation time exceeds the
length of a single series, while being shorter than duration of 
the observational compaign. In this case, processing a short series of frames
will recover noise signature rather than the desired real astrometric signal.
In next section we described results of a statistical analysis of 
${\mathcal N}_{im}$  based on measured residuals $u_{im}$.

\begin{figure}[htb]
{\includegraphics*[bb = 53 49 301 223, width=8.7cm,height=6.3cm]{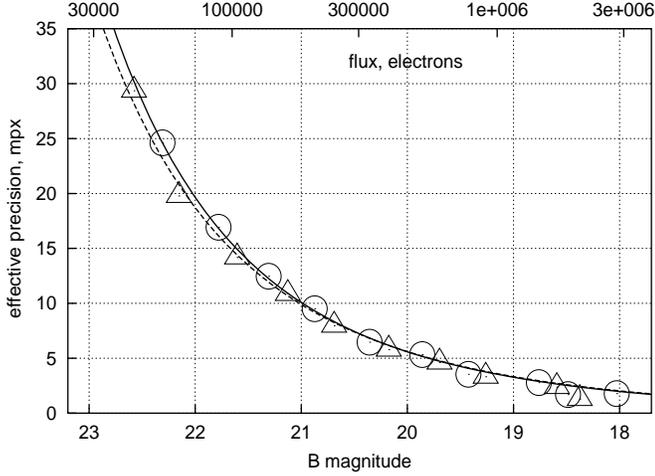}}
\caption { Effective precision  of a single  measurement $\bar{\sigma}$
(mean for $x$ and $y$ axes) for 2000 (circles) and 2002-2003 (triangles) 
 epochs as a function of $B$.
Centroiding error $\bar{\varepsilon}$ 
at average seeing and background noise is shown by solid (first epoch) 
and dashed (second epoch) lines. A scale is 1~mpx=100~$\mu$as.
}
\label{err}
\end{figure}
%%%%%%%%%%%%%%%%%%%%%%%%%%%%%%

The variance  $\sigma_{im}^2$ of the noise ${\mathcal N}_{im}$ is
the effective variance of a single measurement. Obviously,
$\sigma_{im}^2 \geq \varepsilon_{im}^2+ \sigma_{\rm{at}}^2$.
The size of $\sigma_{im}$ depends on the stellar magnitude $I_i$, 
the seeing $\theta_{im}$, the sky level $b_m$, 
the reduction parameters $k$ and $R$, etc.
In the following, the analysis simplifies by
using the improved expression for $\varepsilon_{im}$
that, as compared to Eq.~(\ref{eq:accur}), more adequately takes
into account  variations of  $\theta_{im}$ and 
$b_{m}$, amounting to about $\pm30$\%, and is
valid in a wide (about 5 mag) range of stellar brightness. 
Considering asymptotic dependences of  $\varepsilon$ on $\theta$ and $b$
given by Irwin (\cite{Irwin}) for bright and faint images, we modified
Eq.~(\ref{eq:accur}) to 
\begin{equation}
\label{eq:accur2}
\varepsilon_{im} = \frac{ \theta_{im}}{2.46\sqrt{I_i}}\left (1+ 
	\alpha_0 I_{i}^{-0.7}
          \frac{\theta_{im}}{\theta_{0}}\sqrt{b_m/b_0}  \right ). 
\end{equation}

The second component in parentheses is significant for faint $B>21.5$
stars only. The $\theta_{im}$, or FWHM value, is available as a PSF model
parameter (Sect.2).
We have found  that the variance of the normalized deviations 
$u_{im}/\varepsilon_{im}$ based on our new relation 
is almost independent from observation conditions.
For that reason, the actual observed value of $\sigma_{im} ^2$  was obtained 
as a systematic correction to $\varepsilon_{im}^2$:
\begin{equation}
\label{eq:sigma}
\sigma_{im} ^2= \varepsilon^2_{im} \left \langle \frac{1}{n_l -2} \sum \limits_m 
		u_{lm}^2/ \varepsilon_{lm}^2 \right \rangle
\end{equation}
where an average was taken over stars of approximately the same magnitude. 
Further averaging of 
$\sigma_{im} ^2$ with respect to $i$ and $m$ (over observations)
yielded the effective variance of a single measurement
$\bar{\sigma} ^2$ at average observational conditions during the considered epoch.
Fig.\ref{err} presents both $\bar{\sigma} $ 
and the mean centroiding error $\bar{\varepsilon}$ 
corresponding to average observing conditions
as a function of $B$. Because  $\bar{\sigma} $ matches well the errors
$\bar{\varepsilon}$ predicted by our numerical simulation,
we conclude that the sum of all errors arising from different sources,
including  $\sigma_{\rm{at}}$, 
is small in comparison to $\varepsilon$, so that, 
approximately,
$\bar{\sigma} \approx \bar{\varepsilon}$. 
This means that the astrometric precision depends only on errors
of the photocentre determination which are random Gaussian and uncorrelated
with time due their nature of arising from photon statistics. The latter property
is very important since it forms the basis for substantial
statistical improvements of the astrometric accuracy
$\Delta (n_i)$ for a sequence of $n_i$ frames 
with the increase of the number of measurements as
\begin{equation}
\label{eq:av}
\begin{array}{lr}
\Delta(n_i) = \bar{\varepsilon} /\sqrt{n_i}; & n_i \leq n_{\rm max}\\
\end{array}
\end{equation}
where $\bar{\varepsilon} $ is the photocenter error at average observation
condition. The estimate (\ref{eq:av}) is approximate since assumes the best
situation with a diagonal covariance matrix of the system (\ref{eq:work}).
The validity of Eq.~(\ref{eq:av}) is limited with respect to the
maximum amount of frames $n_{\rm max}$, at which the value of $\Delta(n_i)$
reaches a floor set by systematic errors.

\section{Testing the VLT temporal astrometric stability}
\subsection{The Allan variance}
\begin{figure}[htb]
\includegraphics*[bb = 86 42 284 289, width=8.7cm,height=10.0cm]{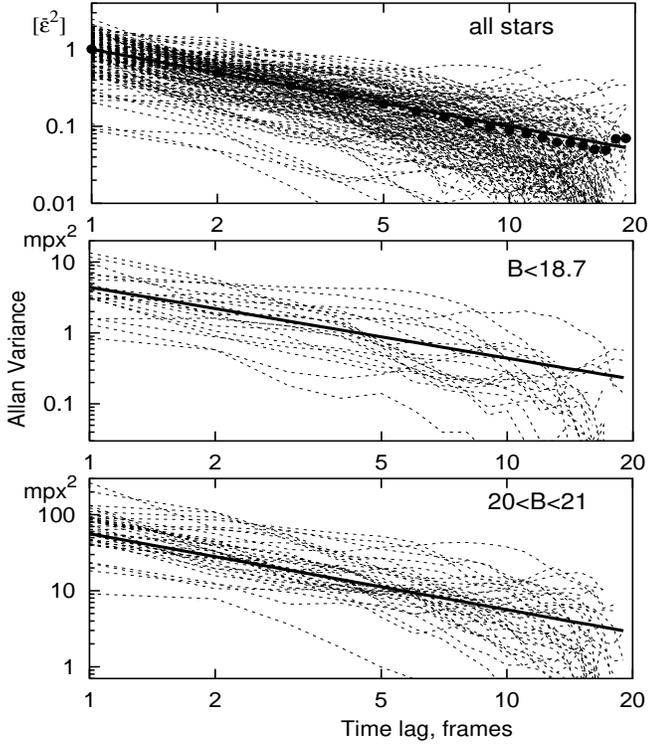}
\caption {Allan variance as a function of time lag 
$\tau$ for individual stars (dashed lines) 
in comparison to the $\tau^{-1}$ law (straight lines)
in narrow magnitude ranges (two lower graphs) and
the normalized variance $\sigma^2_{\rm Al}/ \bar{\varepsilon}^2$
for all  stars (upper plot).  The average
$\langle \sigma^2_{\rm Al}/\bar{\varepsilon}^2 \rangle$  taken over all stars
is shown by black dots. 
The plots refer to $x$ coordinate and 2000 epoch frames.}
\label{allan}
\end{figure}

The simplest
empiric validation of Eq.~(\ref{eq:av}) and illustration of how this law 
works at different $n_i$, is associated with use of Allan variance $\sigma^2_{\rm Al}$
of the residuals $u_{im}$. This quantity 
(Pravdo \& Shaklan \cite{Pravdo96}) is a function of 
time lag $\tau\leq n_i/2$ expressed in frames
\begin{equation}
\label{eq:allaneq}
\sigma^2_{\rm Al}(\tau)=\frac{1}{2(n_i+1-2\tau)}
   \sum \limits_{p=1}^{n_i+1-2\tau} \left [ \tau ^{-1}
       \sum \limits_{q=0}^{\tau -1} (u_{p+q}-u_{p+q+\tau}) \right ] ^2
\end{equation}
$\sigma^2_{\rm Al}(\tau)$ is a precision of the average of $u_{im}$ taken over
$\tau $  frames, and so 
corresponds to $\Delta^2(\tau)$. For uncorrelated
sets of $u_{im}$, the expected  dependence $\sigma^2_{\rm Al} \sim 
\tau ^{-1}$ is consistent with Eq.~(\ref{eq:av}). 
In effort to increase the range of time lags, we performed computations
of $\sigma^2_{\rm Al}$ treating 
all epoch frames as a single series of $n_i$
length. For miscellaneous reasons, $n_i$ varied from about $0.4M$ to $1.0M$.
Results for stars of different $B$ magnitude classes
are shown in  Fig.\ref{allan} (two bottom graphs).
Though, in general, 
plots for separate stars follow the expected law $\tau ^{-1}$ (straight lines
with an ordinate $\bar{\varepsilon}^2$ at $\tau=1$), 
it is difficult
to make any conclusion at $\tau \geq 10$. To smooth statistical
fluctuations and exclude the dependence on the star brightness, 
we  normalized individual $\sigma^2_{\rm Al}$ by $\bar{\varepsilon}^2$ 
(upper plot of Fig.\ref{allan}) and then performed
averaging  of results over all 169 stars. Obtained smooth functions
$\langle \sigma^2_{\rm Al}/\bar{\varepsilon}^2 \rangle$ (black dots) 
indicate no evidence of large systematic error, and 
extend a validity of the approximation 
$\sigma^2_{\rm Al} \sim  \tau ^{-1}$ at least to $\tau \approx 15$.
Hence Eq.(\ref{eq:av}) is valid  to about $n_{\rm max} \approx 15$.  
The asymptotic floor of the Allan
variance at large $\tau$ is under the detection limit of this approach.
We can claim however that with 15 frames, the resulting
precision of the average is about 50$\mu$as  for bright stars. 
In Sect.5.3, we renew this
discussion to derive more exact estimate of $n_{\rm max}$.

%%%%%%%%%%%%%%%%%%%%%%%%%%%%%%

\subsection{Distribution of residuals}

The null hypothesis $H_0$ formally arising from above analysis is that the
${\mathcal N}_{im}$ component in Eqs.~(\ref{eq:work}) represents
an uncorrelated Gaussian noise. The ensuing consequences
of this assumption are 
\begin{itemize}
 \item  the 
    averaging law, Eq.~(\ref{eq:av}), and
 \item the absence of 
    other noise components, including instrumental errors,
   with a noticable magnitude.
\end{itemize}

\begin{figure}[htb]
\begin{tabular}{@{}c@{}c@{}}
{\includegraphics*[bb = 52 60 220 247, width=4.4cm,height=5.3cm]{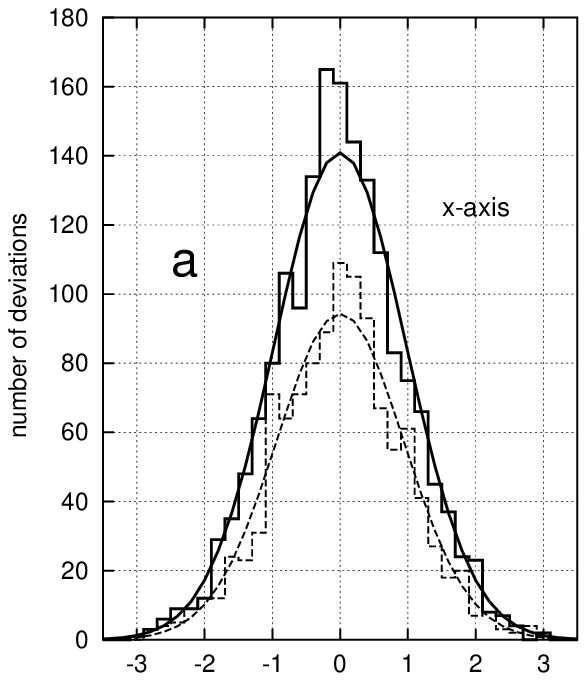}}&
{\includegraphics*[bb = 42 60 220 247, width=4.4cm,height=5.3cm]{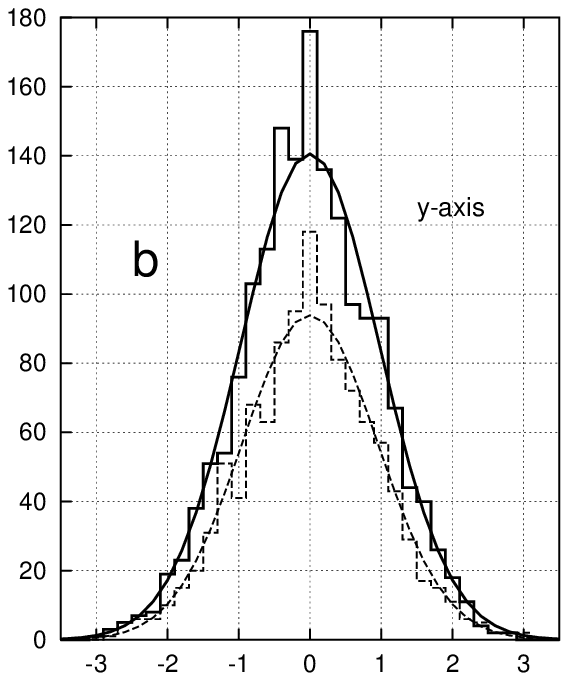}} \\
{\includegraphics*[bb = 48 48 220 247, width=4.4cm,height=6.2cm]{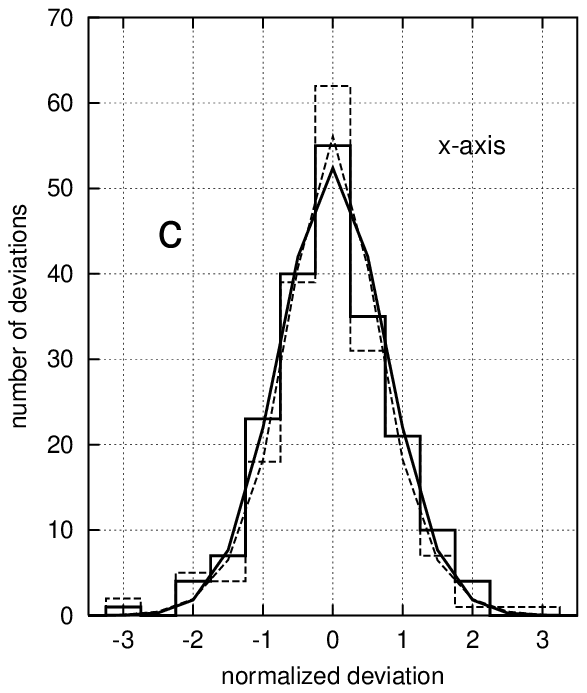}}& 
{\includegraphics*[bb = 36 48 220 247, width=4.4cm,height=6.2cm]{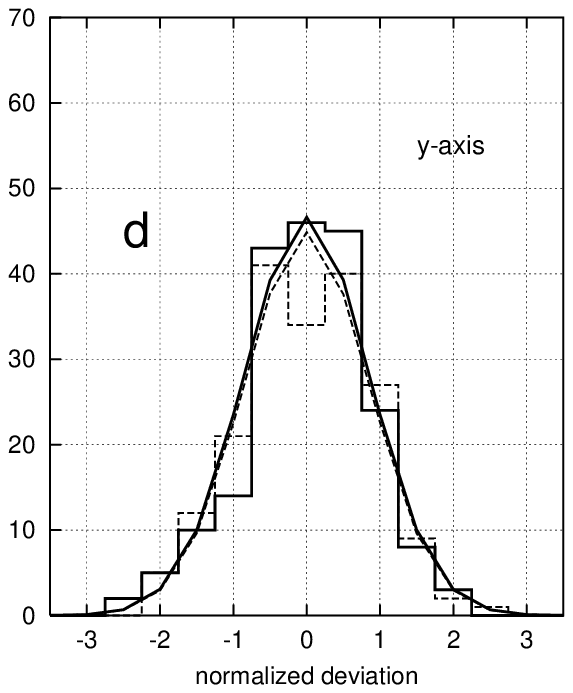}} \\
\end{tabular}
\caption { Histograms of individual normalized residuals
           ${u'}_{im}$ in $x$ ({\bf a}) and        
  in $y$ ( {\bf b}) for
actual (steps) and simulated (smooth lines) observations,
on 2000 (solid) and 2002 epochs (dashed lines);
 the same for normalized night average residuals  $\bar{u}_{ip}/\sigma_{p}$ 
           ({\bf c} and {\bf d}).
          Plots are drawn for bright $B<20.5$ stars.
           }
\label{bin}
\end{figure}

%%%%%%%%%%%%%%%%%%%%%%%%%%%%%%
We applied several statistical tests of our hypothesis.
In the first test, we formed a histogrammic distribution $W$ 
of normalized deviations 
${u'}_{im}=\sqrt{n_i/(n_i-2)}{u}_{im}/\sigma_{im}$ 
and compared it with the distribution $W_0$ of these residuals 
expected when the data fulfill the hypothesis $H_0$. In order
to obtain $W_0$, we performed some numerical simulations of observations with a 
typical spacing of frames on hour angles,
and adding a Gaussian noise with a unit variance
in $x $ and $y $  to the right side of Eqs.~(\ref{eq:work}). 
Observed distributions $W$ were computed 
using all combinations of data sets available 
(2000 and 2003/2003 epoch frames, 
$x$ and $y$ data, with faint, bright, or a whole sample of stars). 
Fig.\ref{bin}a, b present an example of such a histogram $W$ 
for bright stars.
Corresponding simulated distributions $W_0$ are shown 
by smooth curves, for which the same binning as for the computation
of $W$ has been applied. Unlike  ${\mathcal N}_{im}$, the distributions
$W_0$ and $W$ are not Gaussians and have a more strongly peaked shape.
The consistency of observed and model distributions 
by visual inspection is rather good. Numerically, it was estimated 
by $ \chi^2 $ values 
computed from differences between $W$ and $W_0$ in fixed binned intervals
containing at least 5--10 data points. The  $ \chi^2 $ values with
corresponding degrees of freedom (Dof) are listed in Table \ref{x} 
for  faint, bright, or all stars, respectively. 
The $ \chi^2 $-test based on these data does not reveal a deviation of $W$
from $W_0$ at significance levels below 20\%, which does not
allow to reject our hypothesis $H_0$. 
Due to a limited number of data points,
this test is however only sensitive to a central $\pm 2 \sigma$ region of the distribution.

\begin{table}[tbh]
\caption [] {$\chi^2 $ values and Dof characterizing differences 
       between $W$ and $W_0$}
\begin{tabular}{c|ccc}
\hline
\hline
&     faint & bright & all stars \rule{0pt}{11pt}\\ 
\hline
&\multicolumn{3}{c}{ 2000 epoch}  \rule{0pt}{11pt}\\
\hline
$\chi^2$, $x$/$y$ &   20.1 /21.5    &  23.0/22.7 &      24.4/28.6     \\
Dof &   22   &       22        &       24   \\
\hline
&\multicolumn{3}{c}{ 2002/2003 epoch}  \rule{0pt}{11pt}\\
\hline
$\chi^2$, $x$/$y$ &       20.8/11.6    &     25.5/17.2   &     27.8/27.1  \\
Dof &          18      &           20        &       22 \\
\hline
\hline

\end{tabular}
\label{x}
\end{table}

\begin{figure}[htb]
{\includegraphics*[bb = 51 49 301 171, width=8.7cm,height=5.2cm]{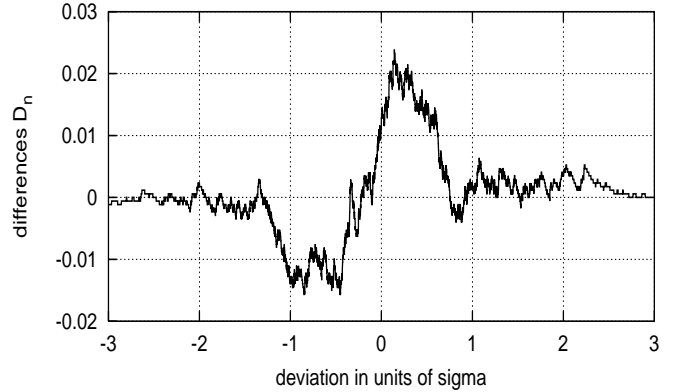}}
\caption {The diffences $D_n$ between the observed and model 
cumulative frequency distributions of normalized residuals
${u'}_{im}$ for 2000 epoch measurements in $y$ (bright stars) 
           }
\label{ks}
\end{figure}

%%%%%%%%%%%%%%%%%%%%%%%%%%%%%%

We applied the more powerful Kolmogorov-Smirnov test to detect a potential
difference between the observed and simulated distributions. For this purpose, we computed
cumulative frequency distributions of observed and model normalized residuals
${u'}_{im}$, and then obtained the difference $D_n$ between two
distributions. A case with largest deviations
for 2000 epoch measurements in $y$ with 1721 data points (bright stars only) 
is shown 
in Fig.\ref{ks} where the differences $D_n$ are plotted as a function
of the normalized deviation. 
The Kolmogorov-Smirnov test based on this plot (in this worst case
max$|D_n|=0.024$) shows 
no significant difference between the model and measured  distributions at 
20\% confidence level. Both tests applied thus do not reject 
hypothesis $H_0$ of
a really normal distribution of ${\mathcal N}_{im}$ at approximately the
same significance levels.

For another test, we considered
nightly averages 
$\bar{u}_{ip}= 
{\sum \limits _{m \in p} }u_{im}\sigma_{im}^{-2} /
{\sum \limits _{m \in p} }\sigma_{im}^{-2}$,
involving all observations of the
$i $-th star within a single $p$-th night. 
Fig.\ref{bin}~c, d show observed and model
frequency  distribution of normalized
residuals $ \bar {u}_{ip}/\sigma_p$, where $\sigma_p $ is the error
of the determination of $ \bar {u}_{ip}$. 
A good agreement of the histograms of the observed statistics (both for bright and faint stars) with their expectations from model simulations,
a symmetry, and the absence of large deviations (normally less than
1\% of $ \bar {u}_{ip}/\sigma_p$ values deviate at about 3 units)
demonstrate excellent stability of the VLT
instrumental system during time spans of a few nights.
Compatability of the observed and model distributions is also confimed
by the Kolmogorov-Smirnov- and  $ \chi^2 $-tests.

\subsection{Integral statistics}
\begin{table}[tbh]
\caption [] {Integral statistics $S_0$ and $S_A$ for bright $B<20.5$,
faint $B>20.5$ stars, and for simulated observations }
\begin{tabular}{r|rrr|rrr}
\hline
\hline
& $S_0^x$ & $S_0^y$ & error  & $S_A^x$  & $S_A^y$  &  error   
			\rule{0pt}{11pt}\\ 
\hline
&\multicolumn{6}{c}{ 2000 epoch}  \rule{0pt}{11pt}\\
\hline
model &  0.997  &  0.985 &  -  &  0.749    &  0.839  &  -  \\
bright &  0.992  & 0.989 & $\pm$0.021    &  0.784    &  0.849  & $\pm$0.064 \\
faint &  0.982  &  0.988  &  $\pm$0.019  &  0.761    &   0.787  & $\pm$0.054  \\
\hline
&\multicolumn{6}{c}{ 2002-2003 epoch}  \rule{0pt}{11pt}\\
\hline
model &  0.994  &  0.973 &  -  &  0.738    &  0.845  &  -  \\
bright &  0.973  & 0.961 & $\pm$0.029    &  0.722    &  0.872  & $\pm$0.067 \\
faint &  0.989  &  0.952  &  $\pm$0.026  &  0.744    &   0.874  & $\pm$0.055  \\
\hline

\end{tabular}
\label{stat}
\end{table}

We also considered integral statistics 
$S_0^2= \langle 
\frac{1}{n_i-4}{\sum \limits _m }\frac{({u}_{im}-\bar{u}_{ip})^2}
{\sigma_{im}^2} \rangle $,
where averages are taken over all stars and nights,
and        $S_A^2= \langle 
 \frac{1}{4}{\sum \limits _p }\bar{u}_{ip}^2/\sigma_{p}^2 \rangle $,
where an average is taken over all stars.
These two statistics are either sensitive to the variation of
systematic errors within a given night ($S_0$) or
from one night to another ($S_A$), and
therefore are a powerful indicator of their presence.
The quantities $S_0 $ and $S_A $, calculated separately for both coordinates
$x $ and $y $, are listed in Table \ref{stat} for each epoch of observations 
together with the respective expected values
$S_0(\rm mod)$ and $S_A(\rm mod) $, obtained from numerical simulations
where random Gaussian noise  ${\mathcal N}_{im}$ has been assumed.
Errors due to a limited number of measurements 
indicate an 80\% confidence interval for
possible deviations of observed values from our model assumption. Observed
statistics match well the model within the error limits. 

The fact that we do
not find an excess in the measured $S_0 $ and $S_A $ values means
that no or very small extra noise except ${\mathcal N}_{im}$ could be 
present in measurements. It is useful to estimate its upper limit.
Suppose that
the observed positions, besides of ${\mathcal N}_{im}$, are
affected by some small systematic error that is constant within a night but
varies between nights with an amplitude peculiar to a certain star. 
The observed value of  $S_A^2(\rm obs)$  then
exceeds its model expectation
$S_A^2(\rm mod) $ by a small amount $S_{\rm sys}^2 $
which is the average signal variance for a given star sample.
These quantities are related by  
$S_A^2(\rm obs) = (\chi^2_{\rm Dof}/{\rm Dof})S_A^2(\rm mod)  
+ S_{\rm sys}^2 $. For statistical reasons, it is probable that
the sampled value $S_A^2(\rm obs)$ 
is equal or below of its mathematical
expectation  $S_A^2(\rm mod) +  S_{\rm sys}^2 $. 
The largest value of $S_{\rm sys}^2 $, 
according to Table~\ref{stat},  can exist in the
$x$-axis measurements of bright stars (2000 epoch), 
for which we found $S_{\rm sys}^2 / S_A^2(\rm mod) 
\leq 0.19$ at 20\% confidence level. This is a ratio of a systematic 
to random error component for a single 
night series represented, in average,  by 8 frames.
In other words, the bias related to systematic errors is equal to
a random component expected in positions averaged over $8/0.19=40$
frames. This is an estimate of $n_{\rm max}$ introduced in
Eq.(\ref{eq:av}),  more precise, 
as compared to that obtained in Sect.5.1. 
For $y$-coordinate, we obtain  $n_{\rm max}=65$.
Frames of 2002 yield  $n_{\rm max}=120$ and 50 
for the $x$ and $y$ axes
correspondingly. With  $n_{\rm max}=50$ adopted as a reliable limit,
precision of a night series is about 30~$\mu$as.

\subsection{Correlations}

\begin{figure}[htb]
\begin{tabular}{@{}c@{}c@{}}
{\includegraphics*[bb = 52 48 223 169, width=4.9cm,height=4.5cm]{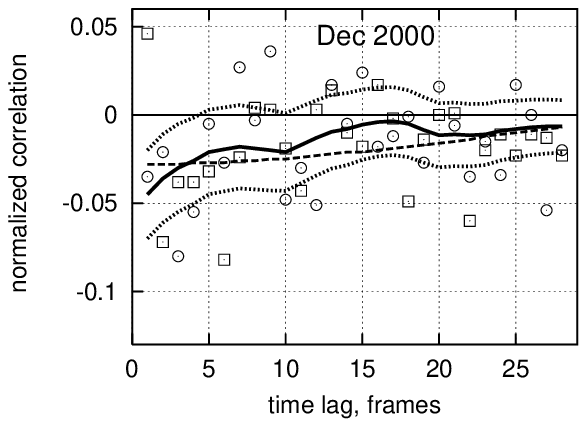}}&
{\includegraphics*[bb = 87 48 223 169, width=3.9cm,height=4.5cm]{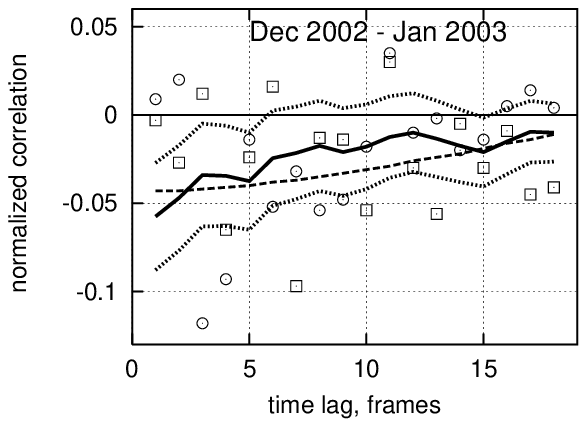}}
\end{tabular}
\caption { Observed autocorrelation function in $x$ (circles) and in $y$
(squares) as compared to the model function (solid line) with a 1-$\sigma $
tolerance limits (two dotted lines) and approximation (\ref{eq:c}) 
(dashed line)} 
\label{cor}             
\end{figure}

A final important test was carried out to investigate correlations in 
${\mathcal N}_{im}$ which can negatively affect the averaging law
(\ref{eq:av}). For that purpose, we computed the 
autocorrelation function of normalized residuals 
\begin{equation}
\label{eq:cm}
{r}(\tau)= H(\tau) \langle (u_{im_1}/\sigma_{im_1})(u_{im_2}/
	{\sigma_{im_2}})  \rangle
\end{equation}
of the argument  $\tau=m_2-m_1$, which is equal to the difference of frame indexes and
used as a time lag.
Here, $H(\tau)=1-\tau /M$ is the weight function 
that compensates  the increase of statistical variations at large
$\tau$. Averaging is performed over the images of all stars.
A corresponding computation of the expected  correlation function 
$\hat{r}(\tau)$
was carried out with the model observations containing white noise
with a unit variance.
The observed ${r}(\tau)$ values computed for both coordinates $x$ and $y$ for bright 
$B<20.5$~mag stars are shown in Fig.\ref{cor} along with the model
function $\hat{r}(\tau)$ (mean for $x$ and $y$) 
and 1-$\sigma$ 
limits for statistical scatter of individual deviations. 
Oscillations in $\hat{r}(\tau)$ plot inherit a compound four-night structure
of the series.
Observed data in general follow the
expected dependence, though with few isolated deviations over 2 sigma.

%%%%%%%%%%%%%%%%%%%%%%%%%%%%%%

Both $r$ and $\hat{r}$ are not identically zero as it could be expected
for uncorrelated measurements. The observed systematic negative bias
of correlations, however, is not due to instrumental errors
but has a theoretical background (Lazorenko \cite{Lazorenko97}).
In order to see this, consider Eqs.~(\ref{eq:work}) for the $i$-th star.
When the number of frames $M$ is large and the
distribution of frames with hour angle is sufficiently random, 
all parameters in the system given by Eq.~(\ref{eq:work}) become 
uncorrelated. Then  $\xi_i^x$ 
is approximately equal to the average of
$(\lambda_{im}^x-h_{im}^x +h_{i0}^x
+{\mathcal N}_{im})$ taken with respect to the index $m$.
In this approximation, one finds 
$u_{im}^x = (\lambda_{im}^x-h_{im}^x +h_{i0}^x
+{\mathcal N}_{im}) -\xi_i^x $.  
The term given in parentheses  formally represents 
a solid time series  of the ''time'' argument $m$ with
$m=1 \ldots M$ measurements containing noise. With this approach, 
the $u_{im}^x$ values  are the remainders of a time series 
after subtraction of the fitting polynomial (namely a constant $\xi_{i}^x$
in our case). The bias of the autocorrelation
function shape due to subtraction of the
best-fitting polynomial for measurements limited in time 
was also previously studied by Lazorenko (\cite{Lazorenko97}).
The autocorrelation function $r'$ of series
remainders takes a particularly simple form
if the power spectrum of measurement errors resembles  
a white noise (which is the case) with a unit
variance and constant rate of data sampling. 
In this case, one obtains (Bakhonski et al. \cite{Bakhonski})
\begin{equation}
\label{eq:c}
r'(\tau)= -\frac{2 \varsigma}{M} \mbox{sinc}(2\pi \varsigma \tau /M)
\end{equation}
where $\varsigma=0.56$,
if the subtracted polynomial is presented by a constant,
and $\mbox{sinc}(z)=\sin (z)/z$.
One can see that $r' \approx 0$ for large $M$ only (long series)
while for small $M$ (limited number of sampled data) 
its value is absolutely large and negative.
Due to simplification of our considerations, the
measured and model functions are not exactly fitted by $r'$, but good enough 
however to explain the 
statistical origin of the observed negative bias (Fig.~\ref{cor}).
We conclude that no correlation is present in ${\mathcal N}_{im}$ at
scales of several frames.

\begin{table}[tbh]
\caption [] {Observed $\bar {r}$ and model $\bar {r}_{\rm {mod}}$
normalized correlations.}
\begin{tabular}{r|rr|rr|r}
\hline
\hline
$\tau '$ & $\bar{r}^x$ & $\bar{r}^x_{\rm{mod}}$  & $\bar{r}^y$  &
$\bar{r}^y_{\rm{mod}}$   & 1-sigma error   
			\rule{0pt}{11pt}\\ 
\hline
&\multicolumn{5}{c}{ 2000 epoch}  \rule{0pt}{11pt}\\
\hline
1 &  -0.17  &  -0.18 &  -0.22  &  -0.24    &  $\pm0.05$  \\
2 &  -0.17  &  -0.10 &  -0.20  &  -0.24    &  $\pm0.07$  \\
3 &  -0.42  &  -0.38 &  -0.25  &  -0.23    &  $\pm0.09$  \\
\hline
&\multicolumn{5}{c}{ 2002-2003 epoch}  \rule{0pt}{11pt}\\
\hline
1 &  -0.15  &  -0.14 &  -0.26  &  -0.25    &  $\pm0.06$  \\
2 &  -0.19  &  -0.20 &  -0.32  &  -0.23    &  $\pm0.07$  \\
3 &  -0.35  &  -0.26 &  -0.13  &  -0.19    &  $\pm0.10$  \\

\hline

\end{tabular}
\label{r}
\end{table}

Correlation at time scales of few days were studied
using normalized nightly-average residuals $\bar{u}_{ip}/ \sigma_{p} $.
We computed correlations
\begin{equation}
\label{eq:ca}
\bar{r}(\tau ')= \langle (\bar{u}_{ip_1}/ \sigma_{p_1})
                  (\bar{u}_{ip_2}/ \sigma_{p_2})
             \rangle
\end{equation}
as a function of the time lag $\tau '=p_2 -p_1$  between two nights
with indexes $p_1$ and $p_2$, expressed in days.
The values of $\bar{r}$ for the observed data and corresponding
model value $\bar{r}_{\rm{mod}}$ are given in Table \ref{r}. 
Apparently, there is a strong negative bias in $\bar{r}$ 
due to a very limited number of data points in our 4-night series.
This bias, however, is well-modelled
using uncorrelated noise ${\mathcal N}_{im}$, which
confirms previous conclusions that there are no traces of
instrumental signature in VLT observations at time scales of few nights,
and supports the validity of the 
averaging law (\ref{eq:av}).

%%%%%%%%%%%%%%%%%%%%%%%%%%%%
\subsection{Two-year stability}

\begin{figure}[htb]
{\includegraphics*[bb = 52 49 267 247, width=8.7cm,height=7.3cm]{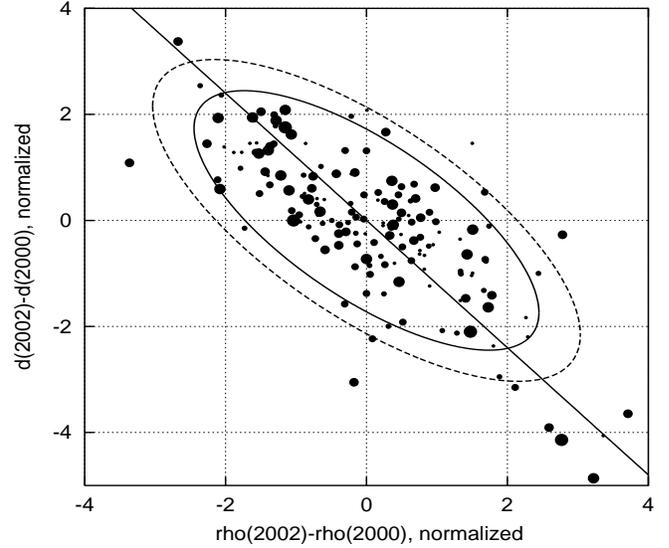}}
\caption { Difference in the
 {\em chromatic} 
 parameters  
$\rho_i$ and $d_i$ between
the two epochs, normalized by the error of this
difference computation (dots). Ellipses are drawn at
1\% (dashed line) and 5\% (solid) significance level.
A straight line shows the direction of point 
displacements due to the actual change of star colours.}
\label{drd}             
\end{figure}

%%%%%%%%%%%%%%%%%%%%%%%%%%%%%%
Given only two epoch measurements, no conclusions on 
a long-term astrometric stability of the VLT can be made based on positional
information. Considering extreme importance of this problem, we present here
indirect analysis of this issue
based, instead of positions,  on a comparison 
of {\em chromatic} $\rho_i $ and $d_i $  parameters 
computed from independent runs of 
2000 and 2002/2003 epochs. 
If one assumes constant colours for the
observed stars, these parameters should
coincide within error bars. 
A change of the VLT characteristics, depending on stellar colours, 
should result in an additional difference. 
Fig.\ref{drd} shows normalized differences between
$  \rho_i $ and $  d_i $ between the two
epochs, where ellipses refer to significance levels of 1\% or 5\%, respectively.
Some points represent outliers, testifying the presence of systematic errors.
The largest 
residuals for bright stars in the right bottom however
occur along a line  $ \rho_i =- 1.2 d_i$ that is peculiar
to the distribution of chromatic parameters shown in Fig.~\ref{rhod}. 
It is therefore very likely
that these deviations are caused by a change in stellar colour. 
Some other large deviations appear to be related to stars in outer regions of the
observed field where the method seems to give biased results. Except for these outliers,
the distribution of points in Fig.~\ref{drd} resembles a Gaussian,
which is an indicator of a rather
good astrometric stability of the VLT even at very long time scales.
It should be stressed that above test is insensitive to non-chromatic type
of errors and therefore is only indicative.

\section{Conclusion}

Our results reveal an
exceptionally good astrometric performance of the VLT and
its camera FORS1. Thus, the precision  $\bar{\varepsilon}$ 
of the position of a bright star reaches 200~$\mu$as 
for a single measurement, which fairly well corresponds to the former 
300~$\mu$as estimate obtained for FORS2 camera
(Paper I). Both cameras thus give similarly good results.
The term ''bright star'' refers to unsaturated images containing
1--3$ \cdot 10^6$ electrons and may correspond to different stellar magnitudes 
depending on the exposure, filter, and seeing.
Most importantly, we found negligibly small systematic
errors of instrumental and other origin. In fact, no traces of these
'dangerous' errors were found at time intervals up to 4 days.
Due to this fact, the precision  for a  frame series  
improves as $\bar{\varepsilon} /\sqrt{n_i}$, and at  
${n_i}= 50$ reaches 
30~$\mu$as for FORS1 and 40~$\mu$as for FORS2.  With use of FORS2, 
the observation time needed to obtain this accuracy
is about 1 hour if short 15~s exposures and a $2\times 2$ binning 
of pixel reading is used.
We conclude that the VLT with cameras FORS1/2, due to its enormous
collecting light power, fine optical performance, and 
effective averaging of wave-front distortions over a large aperture, 
is a powerful instrument that can be used efficiently for
high-precision astrometric observations of short-term events, 
in particular, of planetary microlensing.

%\bibliograhystyle{aa}
%\bibliograhy{fors_bib.tex}

\begin{acknowledgements}
 We would like to thank Dr. E.Jehin for his helpful comments on the
 details of LADC operation.
\end{acknowledgements}

%\bibliograhystyle{aa}

\end{document}